# Interface and Thermophysical Properties of R32 Refrigerant

Abibat Adekoya-Olowofela[1,2], Sukriti Manna[1,2]†, Subramanian KRS Sankaranarayanan[1,2, *]

[1]Center for Nanoscale Materials, Argonne National Laboratory,
Lemont, Illinois 60439, United States

[2]Department of Mechanical and Industrial Engineering, University of Illinois,
Chicago, Illinois 60607, United States

*Corresponding Authors*: †smanna@uic.edu
*skrssank@uic.edu

*Abstract*

*Driven by the urgent demand for efficient cooling in microelectronics and advanced thermal management systems, difluoromethane (R32/$CH_2F_2$) has emerged as a promising candidate owing to its favorable thermophysical properties, including high heat transfer efficiency and low viscosity. While bulk properties such as density, viscosity, and thermal conductivity have been widely studied, interfacial properties including surface tension and interfacial thickness remain comparatively underexplored, despite their importance in phase-transition dynamics. Here, we perform molecular dynamics (MD) simulations from 180 K-300 K using an optimized transferable force field for fluoropropenes with enhanced electrostatics to assess both bulk and interfacial behavior of R32. Simulations reproduced density within ± 2.1%, viscosity within 3.05%, and thermal conductivity within 7.41% of NIST reference data. Heat capacities ($C_p$ and $C_v$) were predicted within 5%. For interfacial properties, surface tension trends were reproduced within 15.8% deviation, and the vapor-liquid coexistence curve closely matched reference data, yielding a critical temperature of 345.7 K (1.6% deviation) and a critical density of 0.397 g/cm³ (6.4% deviation). Importantly, the vapor-liquid interface exhibited strong temperature dependence, with interfacial thickness increasing by 290% between 180 K and 290 K. These validated results provide predictive molecular-level insights, particularly for interfacial properties that remain less characterized. By reducing property prediction errors in key parameters such as critical temperature, this work provides reliable inputs for heat-exchanger and system models. Such correlations can support optimized component sizing, improved performance, and reduced refrigerant charge. Beyond R32, the methodology offers a transferable framework for blended and next-generation low-GWP refrigerants, contributing to sustainable thermal management aligned with the 2027 EU F-Gas regulation and 2030 Kigali Amendment.*

## I. Introduction

High-performance microelectronics and data-intensive computing platforms face growing challenges in dissipating heat as device dimensions shrink and power densities rise [1–4]. Traditional heat transfer fluids often lack the thermophysical balance needed for compact, high-efficiency cooling architectures [5–9]. Difluoromethane (R32, $CH_2F_2$) has emerged as a promising working fluid for advanced thermal management, offering high heat transfer efficiency, relatively low viscosity, and good dielectric stability - making it well-suited for applications such as two-phase immersion cooling, microchannel heat exchangers, and chip-level heat removal. In addition to its favorable performance characteristics, R32 also carries a comparatively low global warming potential (GWP = 677), providing an environmentally advantageous alternative to conventional high-GWP fluids while meeting the stringent thermal demands of next-generation microelectronics. Beyond its environmental advantages, R32 offers excellent thermodynamic behavior, including higher volumetric cooling capacity, superior heat transfer, and improved energy efficiency, making it an attractive option for low-temperature cooling across a broad range of energy applications [10–12].

A detailed analysis of the thermophysical properties of R32, such as density, viscosity, heat capacity, thermal conductivity, surface tension, interfacial thickness, and vapor-liquid equilibrium (VLE), is important for RAC machine design and efficiency. These properties directly influence refrigeration systems' flow characteristics, boiling, and condensation processes, ultimately determining heating, ventilation, air conditioning, and refrigeration (HVACR) systems' performance [13–17]. For example, both boiling and condensation processes are impacted by surface tension; a drop in surface tension enables easier nucleate boiling and increased heat transfer [18,19]. Molecular dynamics (MD) simulations have proven to be powerful tools for predicting bulk properties such as density, viscosity, and thermal conductivity of refrigerants under varying conditions [20–23]. However, interfacial properties such as surface tension and interfacial thickness remain comparatively underexplored, despite their critical role in phase-transition dynamics. Moreover, discrepancies in accurately predicting R32's critical temperature and VLE behavior highlight the need for further investigation [24].

This work addresses these gaps through a comprehensive MD-based study of R32 refrigerant spanning 180 K-300 K, covering a wide range of operational conditions relevant to RAC systems. We use an enhanced force-field parameterization with explicit bond stretching and bending terms, alongside larger simulation system sizes (up to 2,000 molecules for bulk and slab geometries with extended vacuum spacing), thereby improving interfacial property accuracy over prior MD studies. Our primary objective is to establish VLE behavior to clarify phase transitions and critical properties, while also examining interfacial properties to evaluate phase separation and molecular-scale dynamics. In addition, temperature-dependent bulk properties such as density, viscosity, thermal conductivity, and heat capacity are investigated to provide deeper insight into R32's thermodynamic behavior. By linking molecular-scale understanding with practical design, this work aims to provide correlations for viscosity, thermal conductivity, and surface tension that can be integrated into heat-exchanger models. Such correlations will enable manufacturers to optimize component sizing, enhance coefficients of performance, and reduce refrigerant charge, ultimately lowering energy consumption in residential and commercial HVAC equipment. In doing so, our study not only advances scientific knowledge of the properties and atomistic scale behavior of R32 but also delivers practical insights to engineers, modelers, and policymakers working toward more efficient, climate-friendly refrigeration and air-conditioning technologies.

## II. Computational Details and Methodology

We perform molecular dynamics (MD) simulations to examine the interfacial and thermophysical properties of pure R32 refrigerant using the LAMMPS software package [25]. Molecular interactions were described with the Rabbe *et al*. force field, which incorporates bond stretching, angle bending, and non-bonded interactions [26]. These terms preserved molecular geometry and flexibility while capturing dispersion and electrostatics, providing a faithful and transferable description of R32.

**Table 1** summarizes the specific parameters for bond stretching, angle bending, and non-bonded interactions, while Lorentz-Berthelot mixing rules estimate cross-interactions between dissimilar atoms [27]. The molecular geometry and system configurations were designed to accurately capture both bulk and interfacial behavior of R32. **Figure 1a** shows the molecular geometry of R32 used in this study. For the bulk configuration, 2000 R32 molecules were randomly placed using Packmol [28] in a cubic simulation box of dimensions 5 nm × 5 nm × 5 nm (**Figure 1b**). Periodic boundary conditions were applied in all three dimensions to maintain a bulk phase free of surface effects. To study interfacial properties such as surface tension and interfacial thickness, a modified configuration was created. The simulation box was expanded in the z-direction to 5 nm × 5 nm × 20 nm, with 10 nm of vacuum introduced above and below the refrigerant slab, as illustrated in **Figure 1c** [29].

**Table 1. Force field parameters for R32 as reported by Rabbe et al.** [26]

| Atom | ε (Kcal/mol) | σ (Å) | q (e) |
|---|---|---|---|
| C | 0.1085 | 3.15 | 0.43960 |
| H | 0.0157 | 2.2293 | 0.04158 |
| F | 0.0874 | 2.94 | -0.26138 |
| | Bond | $k_d$ (kJ mol$^{-1}$ Å$^{-2}$) | $r_0$ (Å) |
| | C-F | 1544.61 | 1.369 |
| | C-H | 1472.89 | 1.094 |
| | Angle | $k_\theta$ (kJ mol$^{-1}$ rad$^{-2}$) | $\theta_0$ (deg) |
| | H-C-H | 146.54 | 113.6 |
| | F-C-F | 367.61 | 108.7 |
| | H-C-F | 249.92 | 108.6 |

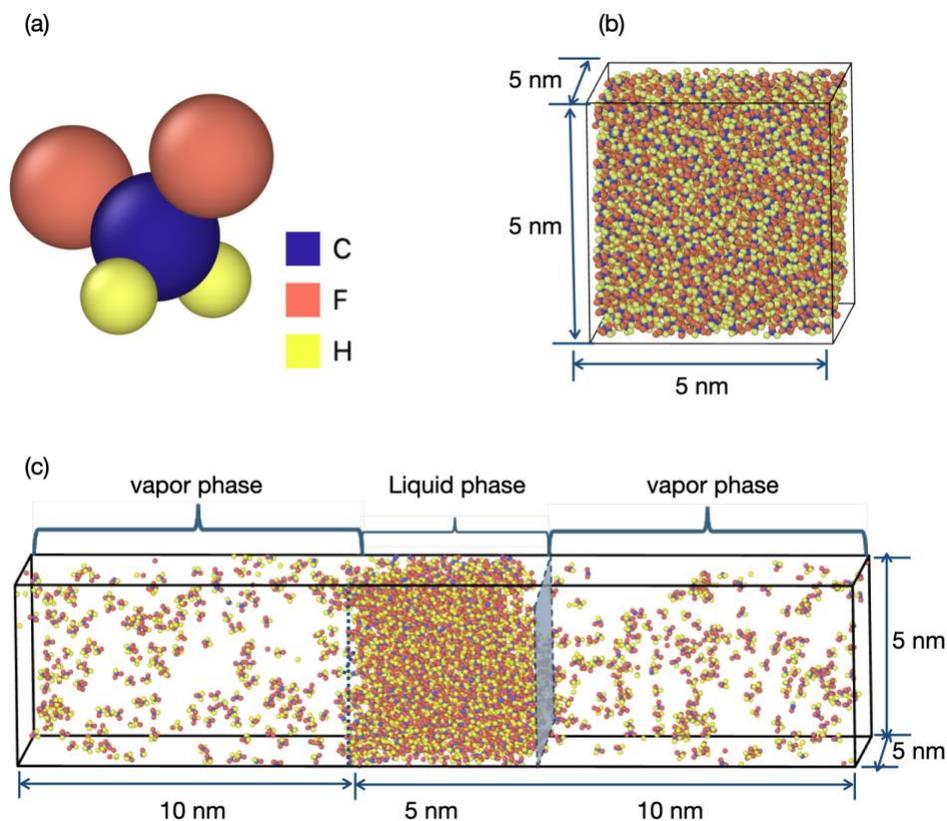

**Figure 1**. **Simulation setup showing the molecular structure of R32 ($CH_2F_2$), initial liquid-phase configuration, and the extended domain setup for the liquid-vapor interface property calculations.** (a) The molecular structure of R32 ($CH_2F_2$), showing a central carbon atom (C, gray) bonded to two hydrogen atoms (H, white) and two fluorine atoms (F, yellow), represents the hydrofluorocarbon used in the simulation. (b) The initial bulk configuration of R32 molecules in the simulation box of size 5 nm × 5 nm × 5 nm. (c) The simulation setup for interface property calculations, with the domain extended to 10 nm × 5 nm × 5 nm.

**Simulation Protocols and Parameters for Thermophysical and Interface Property Analysis of R32**

We performed molecular dynamics (MD) simulations using the Large-scale Atomic/Molecular Massively Parallel Simulator (LAMMPS) [25] to investigate the interfacial and thermophysical properties of the R32 refrigerant across a range of operational temperature regimes. Each system underwent initial energy minimization to obtain a relaxed configuration, ensuring a physically meaningful state. Equilibration was then carried out in the NPT ensemble (constant particle number, pressure, and temperature) at 1 atm and temperatures ranging from 180 to 300 K until both thermal and structural equilibrium were achieved. At each temperature, the protocol included a 1 ns equilibration run followed by a 4 ns production run for property evaluation, with a 1 fs timestep maintained throughout. Intermolecular interactions were modeled with a 12 Å cutoff, long-range Coulombic terms were treated using Ewald summation with a tolerance of $10^{-4}$, and temperature and pressure were regulated using the Nose-Hoover thermostat and barostat[30].

**Density Profile and Interfacial Thickness**

We analyzed density profiles along the z-axis to characterize the liquid-vapor interface of R32. This analysis provides insight into molecular arrangements at the interface and enables estimation of the interfacial thickness. The liquid refrigerant was positioned at the center of the simulation box, with vacuum regions introduced above and below, as illustrated in **Figure 1c**. To obtain the density profile $\rho(z)$, the simulation box was partitioned into 156 bins along the z-axis, and the number of R32 molecules in each bin was used to compute local densities. This procedure yielded continuous density distributions that resolved bulk liquid, interfacial, and vapor regions [31].

The transition zone between the bulk liquid and vapor phases was quantified by fitting the density profile $\rho(z)$ with a hyperbolic tangent function [29]:

$$\rho(z) = \frac{1}{2}(\rho_l + \rho_v) \pm \frac{1}{2}(\rho_l - \rho_v)\tanh(\frac{z-z_0}{d}) \quad (1)$$

where $\rho_l$ and $\rho_v$ are the bulk densities of liquid and vapor phases, respectively, $z_0$ is the midpoint of the interface where the density is halfway between, $\rho_l$ and $\rho_v$ and $d$ represent the average interfacial thickness (**Figure 2a**). Following the 10/90 criterion [29], $d$ was defined as the z-region over which the density profile varied between

$$\rho_v + 0.1(\rho_l - \rho_v) \text{ to } \rho_v + 0.9(\rho_l + \rho_v) \quad [32] \quad (2)$$

After establishing the density profile and interfacial thickness, the coexistence of liquid and vapor phases was examined to construct the vapor–liquid equilibrium (VLE) curve.

**Vapor-Liquid Equilibrium**

We established the phase coexistence behavior of R32 by simulating liquid and vapor phases coexisting in the same system (**Figure 1a**). This configuration allowed us to directly measure equilibrium densities in the bulk liquid $\rho_l$ and vapor $\rho_v$ regions across the studied temperature range. We used these densities to construct the vapor-liquid coexistence curve.

Near the critical temperature, the density difference between liquid and vapor phases was described using a scaling law [31, 33]:

$$(\rho_l - \rho_v) = A(T_c - T)^\beta \quad (3)$$

where A is a fitting constant, T is the temperature, $T_c$ is the critical temperature, and β is the critical exponent that characterizes the critical behavior of the phase transition. Additionally, the law of rectilinear diameters [34] was applied to obtain the critical density:

$$\frac{1}{2}(\rho_l + \rho_v) = \rho_c + B(T_c - T) \quad (4)$$

where $\rho_c$ is the critical density, B is a fitting constant, T is the temperature. By combining equations 3 and 4, the critical temperature $T_c$, critical density $\rho_c$, and constants A, B, and β were computed, which together defined the complete VLE curve for R32. This curve provides detailed insight into the phase-transition characteristics of R32, which are essential for applications requiring the stable coexistence of liquid and vapor phases. Following the phase-behavior analysis, we turned to the evaluation of transport properties.

**Thermal Transport Properties: Thermal Conductivity and Viscosity**

We calculated thermal transport properties, namely, thermal conductivity (κ) and viscosity (η), using equilibrium MD simulations according to the Green-Kubo formalism [35,36]. This approach establishes a relationship between the transport coefficients and the time integrals of both heat flux and shear stress autocorrelation functions. Thermal conductivity was determined from the relation:

$$k = \frac{V}{K_B T^2}\int_0^\infty <J_x(0)J_x(t)> dt \quad (5)$$

where $J_x$ is the heat flux vector, $V$ is the system volume, $K_B$ is the Boltzmann constant, and $T$ is the temperature. We equilibrated each state point for 1 ns in the NVT ensemble and then performed a 7 ns NVE simulation to generate heat flux trajectories. The autocorrelation function $<J_x(0)J_x(t)>$ was computed and integrated to estimate κ.

Viscosity was calculated using an equivalent Green-Kubo scheme:

$$\eta = \frac{V}{K_B T} \int_0^\infty <P_{\alpha\beta}(0)P_{\alpha\beta}(t)> dt \tag{6}$$

where $P_{\alpha\beta}$ are the off-diagonal components of the pressure tensor representing shear stress. As with thermal conductivity, each viscosity calculation included a 1 ns NVT equilibration followed by a 7 ns NVE production run. The autocorrelation of the stress tensor was then computed and integrated to yield η. Interfacial properties were subsequently evaluated using the same configuration.

**Surface Tension**

We calculated surface tension (γ) from the interfacial configuration (Figure 1c). The liquid refrigerant slab was placed centrally in the simulation box with 10 Å vacuum layers extending along the z-direction to form a liquid-vapor interface. The system was equilibrated in the NVT ensemble until thermal equilibrium was achieved. Following equilibration, an NVT production run was performed, and pressure tensor components were recorded. Surface tension was computed from the anisotropy of the pressure tensor by integrating the difference between the normal and tangential pressure components across the interfacial region [33,37]:

$$\gamma = \frac{1}{2} \int_{-\infty}^{\infty} (P_N(z) - P_T(z)) dz \tag{7}$$

where $P_N$ and $Pz$ are the normal and tangential components of the pressure tensor across the interface, and z represents the coordinate perpendicular to the interface. The factor 1/2 accounts for the two interfaces present in the simulation box due to periodic boundary conditions. This approach allowed for the measurement of surface tension across different temperatures. Finally, heat capacities were calculated to complete the thermodynamic characterization.

**Heat Capacity Calculations**

We determined the heat capacities at constant volume ($C_v$) and constant pressure ($C_p$) from thermodynamic data obtained through MD simulations in the NVT and NPT ensembles, respectively. A cubic polynomial fit was applied to energy data, enabling estimation of the heat capacities through analytical differentiation [38,39 40]. For $C_v$, simulations were performed in the NVT ensemble to maintain constant system volume and eliminate work contributions from external forces. Temperature fluctuations were used to sample the total energy (E) over a range of temperatures, and the corresponding energy values were extracted from the MD trajectories. For Cp, simulations were carried out in the NPT ensemble, allowing for volume fluctuations at constant pressure. Enthalpy (H) values were recorded across different temperatures and subsequently analyzed. The dependence of the internal energy (E) and enthalpy (H) on temperature was represented using cubic polynomial models:

$$E(T) = b_0 + b_1 T + b_2 T^2 + b_3 T^3 \tag{8}$$

$$H(T) = a_0 + a_1 T + a_2 T^2 + a_3 T^3 \tag{9}$$

where $b_0$, $b_1$, $b_2$, $b_3$ and $a_0$, a1, $a_2$, $a_3$ are polynomial coefficients obtained by fitting the simulation data using the least-squares method. Once the coefficients were determined, heat capacities were calculated as the first derivatives of the polynomials with respect to temperature.

The isochoric heat capacity $C_v$ was thus expressed as [40]:

$$C_v = \frac{\partial E}{\partial T} = b_1 + 2b_2 T + 3b_3 T^2 \tag{10}$$

and the isobaric heat capacity $C_p$ as:

$$C_p = \frac{\partial H}{\partial T} = a_1 + 2a_2 T + 3a_3 T^2 \tag{11}$$

Substituting the polynomial coefficients into these expressions yielded $C_v$ and $C_p$ values at each studied temperature.

### III. Results and Discussion

In this section, we present the molecular dynamics (MD) simulation results for difluoromethane (R32), focusing on interfacial, thermophysical, and transport properties relevant to its performance as a cooling fluid for microelectronics and advanced thermal management systems. Our analysis spans density profiles, interfacial thickness, surface tension, vapor-liquid equilibrium (VLE), transport properties (thermal conductivity and viscosity), and heat capacities. Where possible, our results are benchmarked against experimental data and previous MD studies to assess the accuracy and predictive power of the employed force field.

**Density Profile and Interfacial Thickness**

Before evaluating transport behavior, we first examined the molecular-scale structure of the liquid-vapor interface. Figure 2a shows the density profile of the vapor-liquid interface of R32, which we inspected to assess molecular layering and the extent of phase separation. The resulting profile reveals a sharp high-density plateau in the liquid phase, a well-defined low-density plateau in the vapor phase, and a transition region connecting the two. In **Figure 2b**, we compared the calculated bulk liquid densities with NIST data [41]. Our simulations slightly underestimated density at lower temperatures but progressively converged toward the reference values at higher temperatures, reflecting improved accuracy at elevated conditions.

We also observe pronounced temperature-dependent changes in interfacial thickness. At lower temperatures, strong intermolecular forces generated a distinct and well-separated interface, producing a comparatively thin transition region. At higher temperatures, by contrast, increased molecular kinetic energy combined with weaker cohesive forces promoted greater diffusion of molecules across the interface, leading to a broader interfacial width (**Figure 2c** and **Figure 2d**). This systematic increase in interfacial thickness with temperature is consistent with an independent MD study [42], confirming that our results fall within the expected range of simulation uncertainty.

As we approach the critical region of R32, the density contrast between liquid and vapor diminishes further, and the interface becomes increasingly diffuse. The simultaneous thickening of the transition region and the disappearance of surface tension at criticality allow for enhanced molecular interpenetration and exchange across phases, which is a hallmark of critical behavior.

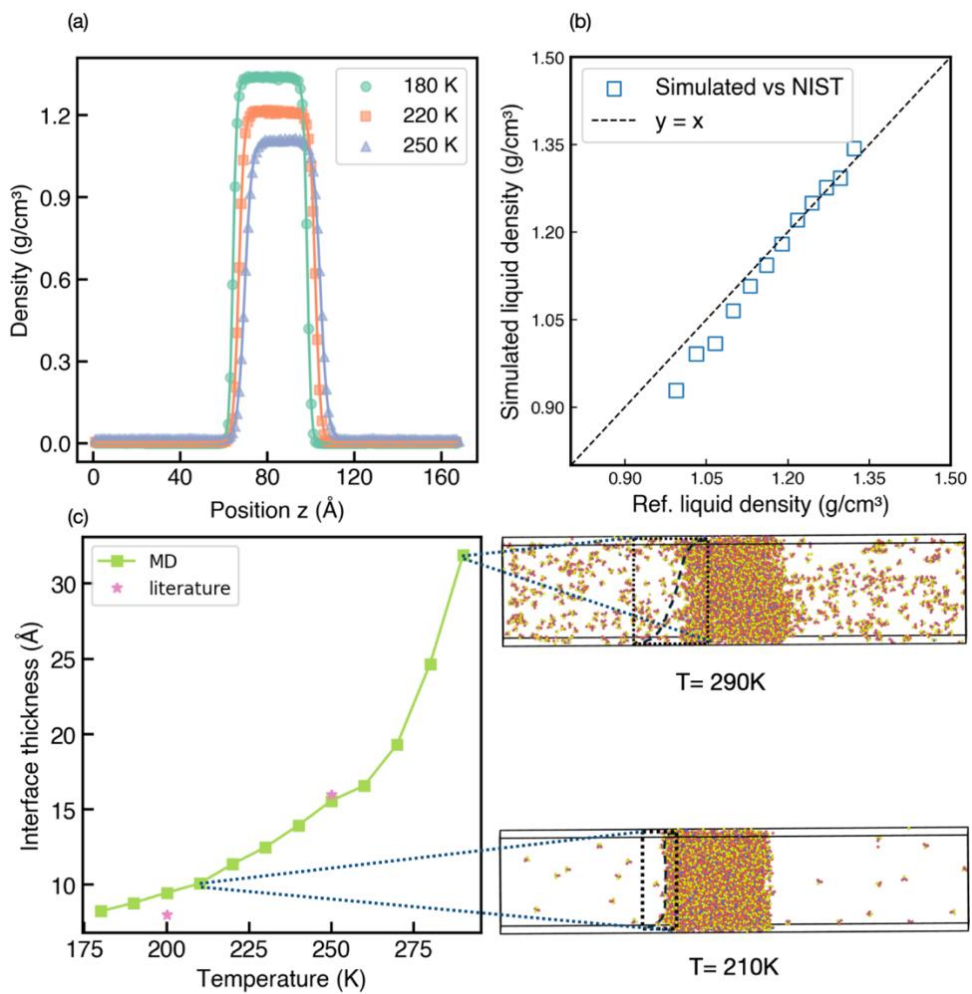

**Figure 2 Density Profiles and Liquid Phase Behavior of R32.** Density profiles of R32 across the liquid-vapor interface at three temperatures: 180 K, 220 K, and 250 K, highlighting the smooth transition zone between the liquid and vapor phases. (a) Liquid density plot against NIST reference values [41], with the diagonal dashed line representing perfect agreement between simulated and referenced data. (use square aspect ratio). Maintain the font size throughout all the subplots. (b) Interfacial thickness variation across various temperatures, illustrating progressive broadening with increasing temperature. Star symbols represent literature data, validating the accuracy of the current study. (c) Snapshots of liquid and vapor phase coexistence for two representative temperatures, 210 K and 220 K, illustrating the width of the interface and the phase separation behavior. A noticeable broadening of the interface at the higher temperature (290 K) is observed, reflecting increased molecular motion and a reduced density contrast between the liquid and vapor phases.

**Surface Tension**

We next calculated the surface tension of R32 and compared it with NIST reference values. Figure 3a presents the temperature dependence of surface tension, showing a clear decreasing trend with increasing temperature. Our simulations produced values that decreased from 0.029 Nm$^{-1}$ at 180 K to 0.0098 N/m at 280 K. These results reproduced the well-known temperature dependence of surface tension and align closely with NIST data. To further validate our model, we compared our results with the independent MD study by Cai et al. [42]. At 200 K, their reported values fell below the reference curve, while our results lay in between, lending confidence to the accuracy of our approach.

We quantified the deviations by calculating the RMSD, which was 0.00264 N/m across the studied range. This value is negligible compared to the absolute surface tension and confirms that our simulations reliably capture the behavior. Agreement with reference data was strong below 220 K, with only minor deviations. At higher temperatures (260 -280 K), our model slightly underestimated surface tension, and discrepancies became more noticeable with increasing temperature. However, these deviations remained within the accepted limits of molecular simulations and are most likely attributable to limitations of the force field at elevated temperatures. Overall, the results confirm that the simulation framework is robust and effective in predicting surface tension trends across a wide temperature range.

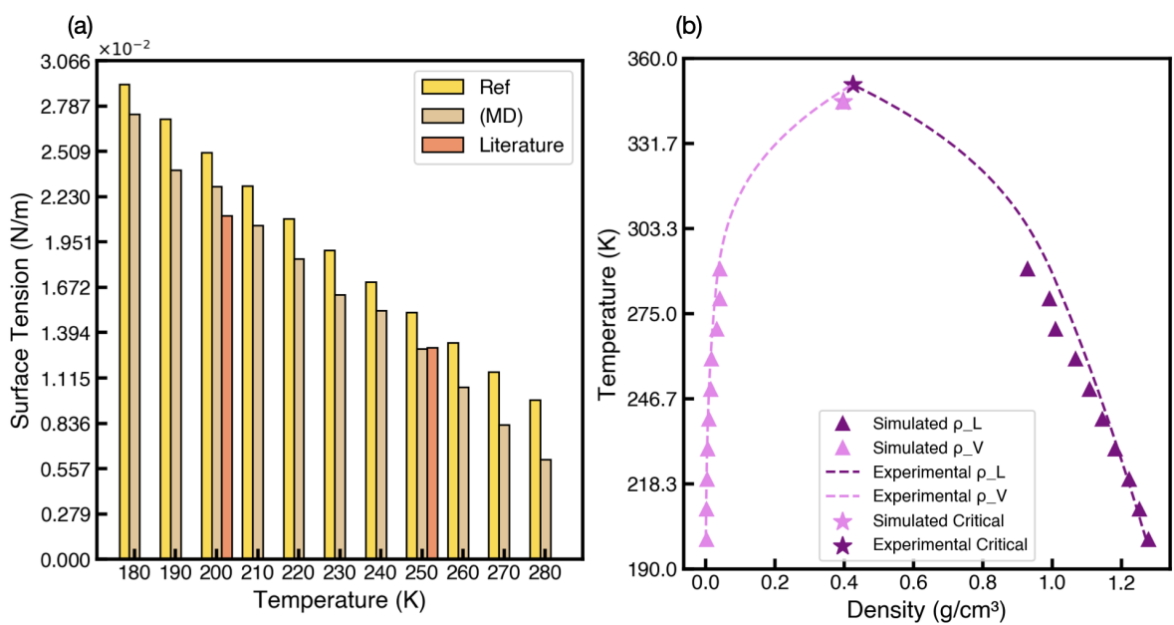

**Figure 3. Surface Tension and Vapor-Liquid Equilibrium Analysis of R32.** (a) Surface tension of R32 as a function of temperature, comparing MD simulation results with reference experimental data and literature values, highlighting the decreasing trend with increasing temperature (b) Vapor-liquid coexistence curve (VLE) displaying liquid and vapor densities across a range of temperatures. Simulated results are compared against reference experimental values, with the critical point marked for both datasets.

**Vapor-Liquid Equilibrium (VLE)**
Building on the interfacial analyses, we then constructed the vapor–liquid coexistence curve of R32 to quantify its two-phase behavior. We constructed the vapor-liquid coexistence curve (VLE) of R32 to describe its two-phase behavior, as shown in **Figure 3b**. With increasing temperature, the liquid density $\rho_l$ decreased, reflecting thermal expansion characteristic of most liquids, while the vapor density $\rho_v$ increased correspondingly. These trends merged at the critical point, where both phases attained equivalent densities.

Our MD simulations produced a VLE curve in close agreement with experimental observations across the studied temperature range. From this analysis, the critical temperature $T_c$ and critical density $\rho_c$ were determined to be 345.7 K and 0.397 g/cm³, respectively, consistent with experimental critical values ($T_c$=351.26 K, $\rho_c$ = 0.424 g/cm³). The liquid and vapor density trends were effectively captured, with discrepancies remaining within an acceptable range. The high level of agreement with experimental observations validates the ability of the simulation model to describe the dual-phase behavior of the refrigerant and its engineering-relevant thermophysical properties.

**Thermal Conductivity and Viscosity**

We extended the analysis from equilibrium and interfacial properties to transport behavior. In particular, we evaluated the thermal conductivity of R32. **Figure 4a** shows the decay of the heat flux autocorrelation function with time at different temperatures, and **Figure 4b** presents the calculated thermal conductivity values across the studied range. Both the simulation output and referenced values depict the expected decrease in thermal conductivity with increasing temperature. The referenced values ranged from 0.22 W/m at 180 K to 0.16 W/m at 250 K. Our calculated results reproduce this behavior with deviations of 7.77% at low temperatures and 6.04% at elevated temperatures. The mean absolute percentage deviation (MAPE) across all studied temperatures was 7.41%. This agreement highlights the effectiveness of the model in capturing the thermal conductivity behavior of R32.

**Figure 4c** illustrates the stress autocorrelation function used to evaluate viscosity, and Figure 4d compares the resulting values with referenced benchmarks. Both datasets show a clear decrease in viscosity with increasing temperature, from approximately 0.0005 Pa·s at 180 K to 0.0002 Pa·s at 250 K. The simulations reproduced this trend with excellent agreement, yielding a MAPE of 3.05% and individual deviations not exceeding 7.41%. This strong agreement confirms the accuracy of the model in capturing the viscosity behavior of R32.

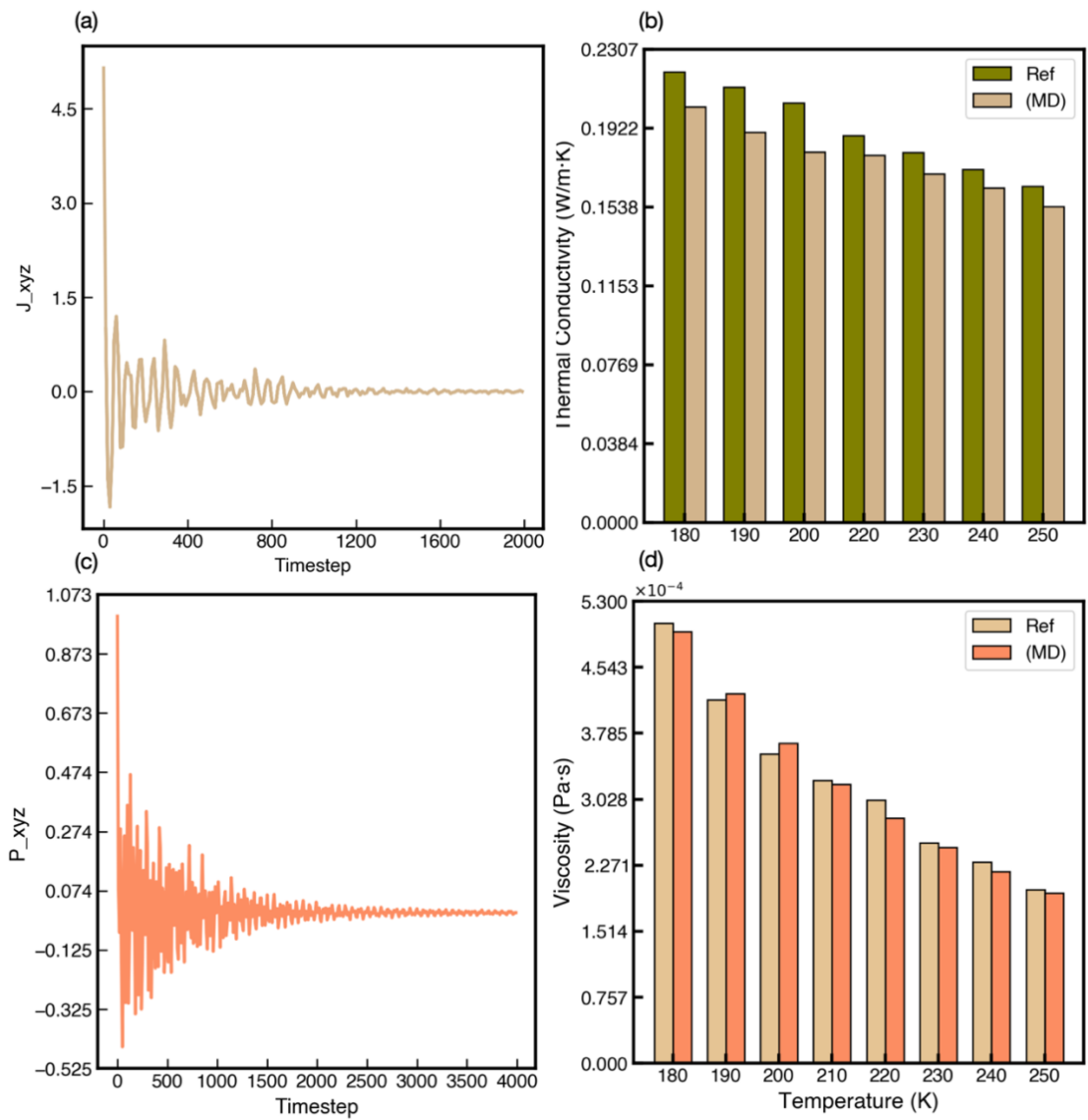

**Figure 4. Thermal Conductivity and Viscosity Analysis of R32 Using Molecular Dynamics Simulations.** (a) Heat flux autocorrelation function used to calculate the thermal conductivity of R32 at various temperatures. (b) Comparison of thermal conductivity values obtained from MD simulations with reference experimental data. (c) Stress tensor autocorrelation function used to compute the viscosity of R32. (d) Comparison of viscosity values obtained from MD simulations with reference experimental data.

**Isochoric ($C_v$) and Isobaric ($C_p$) Heat Capacities**

Finally, to evaluate the energetic and heat storage properties of R32, we computed isochoric ($C_v$) and isobaric ($C_p$) heat capacities under ambient pressure conditions. These quantities are essential for assessing refrigerant energy storage and heat transfer performance, and they directly impact refrigeration cycle efficiency. **Figure 5** compares the simulated $C_v$ and $C_p$ values with NIST reference data across the 230-300 K temperature range, with $C_v$ shown in Figure 5a and $C_p$ in **Figure 5b**. Our simulations reproduced the expected temperature dependence of both properties with strong agreement with reference values. For $C_v$, we observed minor deviations that varied across the temperature range, while $C_p$ closely followed the reference trend throughout. These results confirm that the MD simulations reliably capture heat capacity behavior alongside the other thermophysical properties investigated.

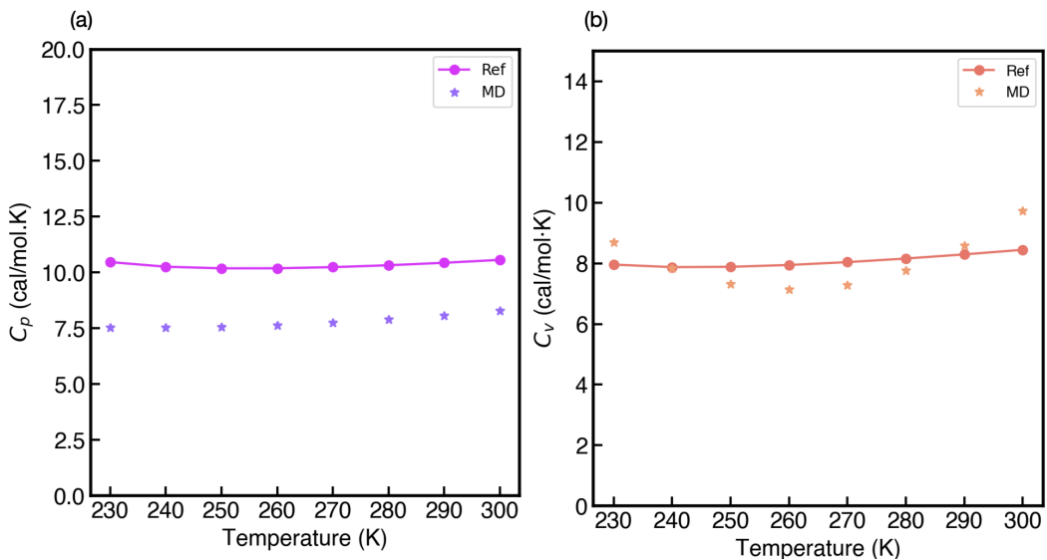

**Figure 5. Variation in the heat capacities of R32 at constant pressure ($C_p$) and constant volume ($C_v$) under a pressure of 1 atm.** (a) shows the heat capacity at constant pressure ($C_p$), derived from molecular dynamics (MD) simulations. The yellow bars represent NIST reference values, while the green bars correspond to MD simulation results. This comparison highlights the agreement between experimental data and the simulation method. (b) presents the heat capacity at constant volume ($C_v$) under the same pressure. The gray bars indicate reference data from the NIST Chem WebBook, [41] and the blue bars represent the MD simulation.

## IV. Conclusion

We investigated the interfacial and thermophysical properties of R32 refrigerant using molecular dynamics simulations. Our results for density, viscosity, thermal conductivity, surface tension, and vapor–liquid equilibrium (VLE) showed excellent agreement with NIST reference data across a wide temperature range. As expected, interfacial width increased with temperature, while surface tension, viscosity, and thermal conductivity decreased. The calculated heat capacities ($C_v$ and $C_p$) agreed closely with experimental values. The VLE curve accurately reproduced the phase behavior of R32, including its critical temperature and density. Together, these results validate the robustness of our methodology and support its predictive power for refrigerant phase transitions. Beyond validation, our study highlights the practical significance of our computational framework. Detailed knowledge of interfacial and thermophysical properties under operational conditions can guide the design of efficient heat exchangers and the development of low-impact refrigerant mixtures. By optimizing for reduced charge volume, minimized pressure drops, and enhanced heat transfer coefficients, our approach enables more efficient heat removal in microelectronics, thereby improving thermal management, reducing hotspot formation, and extending device reliability and lifetime. Moreover, the framework provides a platform for rapid computational screening of next-generation environmentally friendly refrigerants for microelectronics, including low-GWP blends such as R1234ze(E)/R32, R1234yf/R32, R410A, and ternary mixtures. This capability accelerates the discovery of sustainable, high-performance refrigerants aligned with environmental targets and advanced thermal management needs.

**Acknowledgment**

This research was supported by the U.S. Department of Energy (DOE), Office of Science, Office of Basic Energy Sciences, through the Data, Artificial Intelligence, and Machine Learning at DOE Scientific User Facilities program (Award No. 34532, Digital Twins). Portions of the work were performed at the Center for Nanoscale Materials, a DOE Office of Science User Facility funded by the Office of Basic Energy Sciences, under Contracts DE-AC02-06CH11357 and DE-AC05-00OR22725. Computational support was provided by the National Energy Research Scientific Computing Center (NERSC), a DOE Office of Science User Facility under Contract DE-AC02-05CH11231, as well as by the Laboratory Computing Resource Center (LCRC) at Argonne National Laboratory.

*References*
(1) Wang, Z.; Dong, R.; Ye, R.; Singh, S. S. K.; Wu, S.; Chen, C. A Review of Thermal Performance of 3D Stacked Chips. *International Journal of Heat and Mass Transfer* **2024**, *235*, 126212. https://doi.org/10.1016/j.ijheatmasstransfer.2024.126212.
(2) Rahman, S. M. I.; Moghassemi, A.; Arsalan, A.; Timilsina, L.; Chamarthi, P. K.; Papari, B.; Ozkan, G.; Edrington, C. S. Emerging Trends and Challenges in Thermal Management of Power Electronic Converters: A State of the Art Review. *IEEE Access* **2024**, *12*, 50633–50672. https://doi.org/10.1109/ACCESS.2024.3385429.
(3) Dhumal, A. R.; Kulkarni, A. P.; Ambhore, N. H. A Comprehensive Review on Thermal Management of Electronic Devices. *Journal of Engineering and Applied Science* **2023**, *70* (1), 140. https://doi.org/10.1186/s44147-023-00309-2.
(4) Lin, S.-C.; Banerjee, K. Cool Chips: Opportunities and Implications for Power and Thermal Management. *IEEE Transactions on Electron Devices* **2008**, *55* (1), 245–255. https://doi.org/10.1109/TED.2007.911763.
(5) *A comprehensive review on microchannel heat sinks for electronics cooling - IOPscience*. https://iopscience.iop.org/article/10.1088/2631-7990/ad12d4 (accessed 2025-09-22).
(6) *A critical review of traditional and emerging techniques and fluids for electronics cooling - ScienceDirect*. https://www.sciencedirect.com/science/article/abs/pii/S1364032117305944 (accessed 2025-09-22).
(7) Li, C.; Chen, Q.; Hu, Y.; Zou, J.; Su, L.; Shang, Y. Comparative Study of Thermal and Flow Performance of Microchannel Scheme and Multi-Jet Microchannel Scheme Using Liquid Metal and Water as Coolants. *International Journal of Thermal Sciences* **2025**, *211*, 109682. https://doi.org/10.1016/j.ijthermalsci.2025.109682.
(8) Hong, F. J.; Cheng, P.; Ge, H.; Joo, G. T. Conjugate Heat Transfer in Fractal-Shaped Microchannel Network Heat Sink for Integrated Microelectronic Cooling Application. *International Journal of Heat and Mass Transfer* **2007**, *50* (25), 4986–4998. https://doi.org/10.1016/j.ijheatmasstransfer.2007.09.006.
(9) *Review on Coupled Thermo-Hydraulic Performance of Nanofluids and Microchannels*. https://www.mdpi.com/2079-4991/12/22/3979 (accessed 2025-09-22).
(10) van den Bergh, W. J.; Whiting, M.; Theodorakis, P. E.; Everts, M. Nucleate Pool Boiling Bubble Dynamics for R32 and R1234yf on Machined Micro-Structured Surfaces. *International Journal of Thermal Sciences* **2024**, *206*, 109340. https://doi.org/10.1016/j.ijthermalsci.2024.109340.
(11) Vuppaladadiyam, A. K.; Antunes, E.; Vuppaladadiyam, S. S. V.; Baig, Z. T.; Subiantoro, A.; Lei, G.; Leu, S.-Y.; Sarmah, A. K.; Duan, H. Progress in the Development and Use of Refrigerants and Unintended Environmental Consequences. *Science of The Total Environment* **2022**, *823*, 153670. https://doi.org/10.1016/j.scitotenv.2022.153670.

(12) Qian, C.; Yu, B.; Ye, Z.; Shi, J.; Chen, J. Molecular Dynamics Simulation of Interfacial Heat Transfer Characteristics of CO2, R32 and CO2/R32 Binary Zeotropic Mixture on a Smooth Substrate. *International Journal of Refrigeration* **2024**, *157*, 186–198. https://doi.org/10.1016/j.ijrefrig.2023.10.020.

(13) Al Ghafri, S. ZS.; Rowland, D.; Akhfash, M.; Arami-Niya, A.; Khamphasith, M.; Xiong, X.; Tsuji, T.; Tanaka, Y.; Seiki, Y.; May, E. F.; Hughes, T. J. Thermodynamic Properties of Hydrofluoroolefin (R1234yf and R1234ze(E)) Refrigerant Mixtures: Density, Vapour-Liquid Equilibrium, and Heat Capacity Data and Modelling. *International Journal of Refrigeration* **2019**, *98*, 249–260. https://doi.org/10.1016/j.ijrefrig.2018.10.027.

(14) Bobbo, S.; Nicola, G. D.; Zilio, C.; Brown, J. S.; Fedele, L. Low GWP Halocarbon Refrigerants: A Review of Thermophysical Properties. *International Journal of Refrigeration* **2018**, *90*, 181–201. https://doi.org/10.1016/j.ijrefrig.2018.03.027.

(15) *Refrigerant performance evaluation including effects of transport properties and optimized heat exchangers - ScienceDirect*. https://www.sciencedirect.com/science/article/pii/S0140700717302037?via%3Dihub (accessed 2024-12-17).

(16) Di Nicola, G.; Di Nicola, C.; Moglie, M. A New Surface Tension Equation for Refrigerants. *Int J Thermophys* **2013**, *34* (12), 2243–2260. https://doi.org/10.1007/s10765-011-0991-1.

(17) Kondou, C.; Nagata, R.; Nii, N.; Koyama, S.; Higashi, Y. Surface Tension of Low GWP Refrigerants R1243zf, R1234ze(Z), and R1233zd(E). *International Journal of Refrigeration* **2015**, *53*, 80–89. https://doi.org/10.1016/j.ijrefrig.2015.01.005.

(18) Imai, T.; Kawahara, T.; Nonaka, R.; Tomassetti, S.; Okumura, T.; Higashi, Y.; Di Nicola, G.; Kondou, C. Surface Tension Measurement and Molecular Simulation for New Low Global Warming Potential Refrigerants R1132(E) and R1132a. *Journal of Molecular Liquids* **2024**, *407*, 125262. https://doi.org/10.1016/j.molliq.2024.125262.

(19) Mota-Babiloni, A.; Navarro-Esbrí, J.; Makhnatch, P.; Molés, F. Refrigerant R32 as Lower GWP Working Fluid in Residential Air Conditioning Systems in Europe and the USA. *Renewable and Sustainable Energy Reviews* **2017**, *80*, 1031–1042. https://doi.org/10.1016/j.rser.2017.05.216.

(20) *Molecular Simulation for Thermodynamic Properties and Process Modeling of Refrigerants | Journal of Chemical & Engineering Data*. https://pubs.acs.org/doi/10.1021/je500260d (accessed 2024-12-17).

(21) Khan, M.; Wen, J.; Shakoori, M. A.; Zhou, A. Thermophysical Properties and Condensation of R514A through Molecular Dynamics Simulation. *International Journal of Thermofluids* **2023**, *20*, 100436. https://doi.org/10.1016/j.ijft.2023.100436.

(22) Agbodekhe, B.; Marin-Rimoldi, E.; Zhang, Y.; Dowling, A. W.; Maginn, E. J. Assessment and Ranking of Difluoromethane (R32) and Pentafluoroethane (R125) Interatomic Potentials Using Several Thermophysical and Transport Properties Across Multiple State Points. *J. Chem. Eng. Data* **2024**, *69* (2), 427–444. https://doi.org/10.1021/acs.jced.3c00379.

(23) Qin, L.; Shen, L.; Hu, Y.; Zhou, R.; Li, S. Molecule Dynamic Simulation on the Effect of Lubricant on R32 during Condensation. *International Communications in Heat and Mass Transfer* **2024**, *156*, 107663. https://doi.org/10.1016/j.icheatmasstransfer.2024.107663.

(24) *Molecular dynamics of fluoromethane type I hydrates - ScienceDirect*. https://www.sciencedirect.com/science/article/pii/S0167732221014446?via%3Dihub (accessed 2024-12-17).

(25) Thompson, A. P.; Aktulga, H. M.; Berger, R.; Bolintineanu, D. S.; Brown, W. M.; Crozier, P. S.; in 't Veld, P. J.; Kohlmeyer, A.; Moore, S. G.; Nguyen, T. D.; Shan, R.; Stevens, M. J.; Tranchida, J.; Trott, C.; Plimpton, S. J. LAMMPS - a Flexible Simulation Tool for Particle-Based Materials Modeling at the Atomic, Meso, and Continuum Scales. *Computer Physics Communications* **2022**, *271*, 108171. https://doi.org/10.1016/j.cpc.2021.108171.

(26) Raabe, G. Molecular Simulation Studies on the Vapor–Liquid Phase Equilibria of Binary Mixtures of R-1234yf and R-1234ze(E) with R-32 and CO2. *J. Chem. Eng. Data* **2013**, *58* (6), 1867–1873. https://doi.org/10.1021/je4002619.
(27) *Ueber die Anwendung des Satzes vom Virial in der kinetischen Theorie der Gase - Lorentz - 1881 - Annalen der Physik - Wiley Online Library*. https://onlinelibrary.wiley.com/doi/10.1002/andp.18812480110 (accessed 2024-12-17).
(28) *Packmol - Initial configurations for Molecular Dynamics*. https://m3g.github.io/packmol/citation.shtml (accessed 2025-09-11).
(29) Allen, M. P.; Tildesley, D. J. *Computer Simulation of Liquids*, Second edition.; Oxford University Press: Oxford, United Kingdom, 2017.
(30) Evans, D. J.; Holian, B. L. The Nose–Hoover Thermostat. *The Journal of Chemical Physics* **1985**, *83* (8), 4069–4074. https://doi.org/10.1063/1.449071.
(31) Rowlinson, J. S, W., B. *Molecular Theory of Capillarity*, Revised edition.; Oxford University Press: Oxford, 1989.
(32) Muller, E. A.; Ervik, Å.; Mejía, A. A Guide to Computing Interfacial Properties of Fluids from Molecular Simulations [Article v1.0]. *Living Journal of Computational Molecular Science* **2020**, *2* (1), 21385–21385. https://doi.org/10.33011/livecoms.2.1.21385.
(33) Mecke, M.; Winkelmann, J.; Fischer, J. Molecular Dynamics Simulation of the Liquid–Vapor Interface: The Lennard-Jones Fluid. *The Journal of Chemical Physics* **1997**, *107* (21), 9264–9270. https://doi.org/10.1063/1.475217.
(34) J S Rowlinson, F. L. S. *Liquids and Liquid Mixtures*, Third Edition.; 1982.
(35) Kubo, R. Statistical-Mechanical Theory of Irreversible Processes. I. General Theory and Simple Applications to Magnetic and Conduction Problems. *J. Phys. Soc. Jpn.* **1957**, *12* (6), 570–586. https://doi.org/10.1143/JPSJ.12.570.
(36) Green, M. S. Markoff Random Processes and the Statistical Mechanics of Time-Dependent Phenomena. II. Irreversible Processes in Fluids. *The Journal of Chemical Physics* **1954**, *22* (3), 398–413. https://doi.org/10.1063/1.1740082.
(37) Blokhuis, E. M.; Bedeaux, D.; Holcomb, C. D.; Zollweg, J. A. Tail Corrections to the Surface Tension of a Lennard-Jones Liquid-Vapour Interface. *Molecular Physics* **1995**, *85* (3), 665–669. https://doi.org/10.1080/00268979500101371.
(38) Peris, C. S.; Coorey, R. V.; Weerasinghe, S. Molecular Dynamic Simulation of Heat Capacities and Diffusion Coefficients as a Function of Temperature of Triatomic Molecules. *Sri Lanka* **2010**.
(39) Pi, H. L.; Aragones, J. L.; Vega, C.; Noya, E. G.; Abascal, J. L. F.; Gonzalez, M. A.; McBride, C. Anomalies in Water as Obtained from Computer Simulations of the TIP4P/2005 Model: Density Maxima, and Density, Isothermal Compressibility and Heat Capacity Minima. *Molecular Physics* **2009**, *107* (4–6), 365–374. https://doi.org/10.1080/00268970902784926.
(40) *(PDF) Anomalies in water as obtained from computer simulations of the TIP4P/2005 model: Density maxima, and density, isothermal compressibility and heat capacity minima*. https://www.researchgate.net/publication/45853186_Anomalies_in_water_as_obtained_from_computer_simulations_of_the_TIP4P2005_model_Density_maxima_and_density_isothermal_compressibility_and_heat_capacity_minima (accessed 2025-07-07).
(41) Eric W. Lemmon; Bell, I. H.; Marcia L. Huber; Mark O. McLinden. Thermophysical Properties of Fluid Systems in NIST Chemistry WebBook, NIST Standard Reference Database Number 69, Eds. P.J. Linstrom and W.G. Mallard, National Institute of Standards and Technology, Gaithersburg MD, 20899.
(42) Cai, S.; Li, X.; Yu, L.; Zhang, L.; Huo, E. Thermodynamic and Mass Transport Properties of R1234ze(E) and R32 Mixtures at the Liquid-Vapor Interface: A Molecular Dynamics Study. *Journal of Molecular Liquids* **2022**, *365*, 120112. https://doi.org/10.1016/j.molliq.2022.120112.